\documentstyle[preprint,aps,epsf,tighten]{revtex}

\newcommand{\slashed}[1]{\rlap{$#1$}/}
\newcommand{\slashp}{\mbox{$\not \hspace*{-1.10mm} p$}}

\newcommand{\NSt}{{\mbox{\scriptsize\it NS}}}
\newcommand{\St}{{\mbox{\scriptsize\it S}}}
\newcommand{\rip}[1]{\left| {#1} \right\rangle}

\begin{document}

\draft

\preprint{ZTF-01/05}


\title{ 
$\eta$ and $\eta^\prime$ in a coupled Schwinger--Dyson and \\ Bethe--Salpeter 
approach. II. The $\gamma^\star\gamma$ transition form factors
}

\author{Dalibor Kekez}
\address{\footnotesize Rudjer Bo\v{s}kovi\'{c} Institute,
         P.O. Box 180, 10002 Zagreb, Croatia}

\author{Dubravko Klabu\v{c}ar}
\address{\footnotesize Department of Physics, Faculty of Science, \\
        Zagreb University, P.O. Box 331, 10002 Zagreb, Croatia}

\maketitle

\begin{abstract}

\noindent The applications of the consistently coupled Schwinger--Dyson 
and Bethe--Salpeter approach to the $\eta$--$\eta^\prime$ complex are 
extended to the 
two--photon transition form factors of $\eta$ and $\eta^\prime$
for spacelike transferred momenta. 
We compare our predictions with experiment and some other theoretical 
approaches. 

\end{abstract}

{\bf 1.}
In Ref. \cite{KlKe2}, we studied the 
$\eta$--$\eta^\prime$ complex and its $\gamma\gamma$ decays in 
a coupled Schwinger-Dyson and Bethe-Salpeter (SD-BS) approach
(reviewed recently in, e.g., \cite{RW,AlkoferSmekal00,RobertsSchmidt00}).
We obtained the pertinent masses, the pseudoscalar state mixing
angle $\theta$, the results for the axial-current decay
constants of $\eta_8$, $\eta_0$ and of their physical combinations
$\eta$ and $\eta^\prime$, the results for the $\gamma\gamma$-decay
constants of $\eta_0$ and $\eta_8$, for the two--photon decay widths
of $\eta$ and $\eta^\prime$, and for the mixing--independent $R$--ratio
constructed from them. On the other hand, the form factors for the 
transitions $\gamma^\star \gamma \rightarrow \eta$ and 
$\gamma^\star \gamma \rightarrow \eta^\prime$, where $\gamma^\star$ 
denotes an off--shell photon, were {\it not} studied in Ref. \cite{KlKe2}
although a previous reference \cite{KeKl1} of ours 
(see also later Ref. \cite{KeBiKl98}) addressed the 
closely related topic of the $\gamma^\star \gamma \rightarrow \pi^0$ 
transition form factor. Namely, in that paper the pion was treated 
{\it in the chiral (and soft) limit}, which 
made plausible certain simplifications in the 
description of the pseudoscalar quark--antiquark ($q\bar q$) bound state
[see the approximation (\ref{chLimSol}) below],
making the calculation a lot easier.    
Nevertheless, $\eta$ and $\eta^\prime$ contain significant $s\bar s$ 
components, which are rather massive. This obviously makes such 
chiral-limit-based simplifications implausible in a {\it quantitative} 
treatment of the $\gamma^\star \gamma \rightarrow \eta$ and
$\gamma^\star \gamma \rightarrow \eta^\prime$ form factors. 
Having to refrain therefore from the chiral limit simplifications,
while having to deal with the complications due to one off-shell photon,
made Ref. \cite{KlKe2} relegate to a later paper the extension of 
the $\eta, \eta' \to \gamma\gamma$ calculations to the off--shell case. 

Now, however, we are ready to take up the task of studying the off--shell
$\gamma^\star \gamma \rightarrow \eta, \eta^\prime$ amplitudes and 
supplement the results of Ref. \cite{KlKe2} with them, because
our subsequent Ref. \cite{KeKl3} went beyond the chiral and soft
limit approximation when calculating the pion $\gamma^\star \gamma$
transition form factor. In other words, it went beyond the approximation 
where the $q\bar q$ bound state pseudoscalar BS vertex (say, of $\pi^0$)
of the total momentum $p$,
\begin{equation}
\Gamma_{\pi^0}(q,p) \equiv \frac{\lambda^3}{\sqrt{2}} \,
{\rm diag} \left[\, \Gamma_{u\bar u}(q,p), \Gamma_{d\bar d}(q,p), 
\Gamma_{s\bar s}(q,p) \, \right] \, ,
\end{equation}
is approximated by its leading ${\cal O}(p^0)$ piece, 
which depends only on $q$, the relative momentum of the constituents: 
\begin{equation}
\Gamma_{\pi^0}(q,p) 
\approx \Gamma_{\pi^0}(q,0)_{m=0} 
= \gamma_5 \, \lambda^3 \, \frac{B(q^2)_{m=0}}{f_\pi} 
\, .
\label{chLimSol}
\end{equation}
Here $\Gamma_{f\bar f}$ denotes the 
$q\bar q$ bound state vertex for the 
flavor $f(=u,d,s)$,
while $\lambda^a$ denotes the $a$-th ($a=1,...,8$) 
Gell-Mann matrix of the flavor SU(3), with
$\lambda^3$ being the pertinent one for the neutral pion $\pi^0$, and
$B(q^2)$ is the scalar function from the SD solution for the dynamically 
dressed quark propagator $S(q) = [ A(q^2)\slashed{q} - B(q^2) ]^{-1}$.
The subscript $m=0$ indicates the case of vanishing {\it explicit} chiral 
symmetry breaking. This case is quite close to reality in the case
of pions, which are almost massless.  Nevertheless, 
in contradistinction to the SD-BS calculation of the on-shell decay 
$\pi^0 \to \gamma\gamma$,
keeping only ${\cal O}(p^0)$ terms, as in the right-hand side of 
Eq. (\ref{chLimSol}), turned out to be 
rather inadequate for calculation of the $\gamma^\star\gamma\to\pi^0$ 
form factors even in the pion case \cite{KeKl3,KlKe4}. 
Therefore, 
Ref. \cite{KeKl3} used a {\it complete} solution for the BS vertex 
$\Gamma_{\pi^0}(q,p)$ [or equivalently, for the BS amplitude 
$\chi_{\pi^0}(q,p)\equiv S(q+{p}/{2})\Gamma_{\pi^0}(q,p)S(q-{p}/{2})$],
given by the decomposition into four scalar functions 
$\Gamma^{f\bar f}_i(q,p)$
multiplying independent spinor structures, as in Eq. (9) in Ref. \cite{KlKe2}: 
        \begin{equation}
\Gamma_{f\bar f}(q,p)=
        \gamma_5 \left\{\, \Gamma^{f\bar f}_0(q,p) + 
\slashp \, \Gamma^{f\bar f}_1(q,p)
      + \slashed{q} \, \Gamma^{f\bar f}_2(q,p) +
        [\slashp,\slashed{q}]\, \Gamma^{f\bar f}_3(q,p)\, \right\}.
\label{Decomposition}
        \end{equation}
In the isospin limit, which we adopt 
as an excellent approximation, the $u\bar u$ and $d\bar d$ bound states have 
identical BS vertices, $\Gamma_{u\bar u}(q,p)=\Gamma_{d\bar d}(q,p)$.
However, the BS vertex $\Gamma_{s\bar s}(q,p)$, pertaining to the 
much more massive strange quarks, $f=s$, is significantly different
\cite{munczek92,jain93b,KlKe2}.

Using a complete solution (\ref{Decomposition}) in the manner of 
Ref. \cite{KeKl3}, but now also for $f=s$, makes us able to calculate 
adequately the off-shell 
amplitudes $T_{f\bar f}(k^2,{k'}^2)$ for the transitions from 
$\gamma^\star(k) \gamma^{(\star)}(k')$ to ${f\bar f}$ pseudoscalar
of momentum $p= k+k'$, where $k^2\neq 0$ (and possibly also ${k'}^2\neq 0$).  
See Eqs. (27) and (24)-(25) in Ref. \cite{KlKe2} for the 
definition of $T_{f\bar f}(k^2,k^{\prime 2})$ and the explicit expression 
used both for calculating it there for the on--shell case
($k^2={k'}^2=0$), and in the present off--shell application.
The $\gamma^\star \gamma^{(\star)}$ transition amplitudes of the physical 
particles $\eta$ and $\eta'$, denoted, respectively, by $T_\eta(k^2,{k'}^2)$ and 
$T_{\eta^\prime}(k^2,{k'}^2)$, are then obtained as the appropriate mixtures 
of $T_{f\bar f}(k^2,k^{\prime 2})$ -- e.g., as the obvious off--shell
generalization of Eqs. (38)--(39) in Ref. \cite{KlKe2}.                                         
\vspace{3mm}

{\bf 2.}
The mixing scheme used in Ref. \cite{KlKe2} was the octet--singlet one,
where $\eta$ and $\eta^\prime$ are given
through the octet--singlet mixing angle $\theta$ and the SU(3)$_f$ octet 
and singlet isospin zero states $\eta_8$ and $\eta_0$. They are in turn
defined in the $f\bar f$ ($f=u,d,s$) basis by 
        \begin{eqnarray}
        |\eta_8\rangle
        &=&
        \frac{1}{\sqrt{6}} (|u\bar{u}\rangle + |d\bar{d}\rangle
                                            -2 |s\bar{s}\rangle)~,
\label{eta8def}
        \\
        |\eta_0\rangle
        &=&
        \frac{1}{\sqrt{3}} (|u\bar{u}\rangle + |d\bar{d}\rangle
                                             + |s\bar{s}\rangle)~,
\label{eta0def}
        \end{eqnarray}
where it should be noted that the model calculations in Ref. \cite{KlKe2}
and here employ the {\it broken} SU(3)$_f$ with an $s$-quark 
realistically more massive than $u$- and $d$-quarks.

In this paper, nevertheless, we opt to use a mixing scheme different from 
the one in Ref. \cite{KlKe2}, namely the nonstrange ($NS$)--strange ($S$) 
scheme.
In Ref. \cite{KlKe2}, it was essential to explain the successful reproduction of 
the Goldstone character of the SU(3)$_f$ octet state $\eta_8$ and the 
non-Goldstone character of the SU(3)$_f$ singlet state $\eta_0$, since
the U$_A$(1) anomaly causes $\eta' \to \eta_0$ to remain massive even 
though $\eta \to \eta_8$ becomes massless when the chiral limit is taken 
for all three flavors, $m_u, m_d, m_s \to 0$. The important role of  
$\eta_8$ and $\eta_0$, Eqs. (\ref{eta8def}) and (\ref{eta0def}), in the 
discussions in Ref. \cite{KlKe2} made the octet--singlet mixing scheme the most 
convenient one to use there. 
Here, however, it is somewhat more convenient to work in the 
{\it NS}--{\it S} basis $|\eta_\NSt\rangle$ and $|\eta_\St\rangle$, where
        \begin{eqnarray}
 | \eta_\NSt \rangle
        &=&
        \frac{1}{\sqrt{2}} (|u\bar{u}\rangle + |d\bar{d}\rangle)
  = \frac{1}{\sqrt{3}} |\eta_8\rangle + \sqrt{\frac{2}{3}} |\eta_0\rangle~,
\label{etaNSdef}
        \\
 |\eta_\St\rangle &=& |s\bar{s}\rangle 
   = - \sqrt{\frac{2}{3}} |\eta_8\rangle + \frac{1}{\sqrt{3}} |\eta_0\rangle~,
\label{etaSdef}
        \end{eqnarray}                                                                         
and where the {\it NS--S} mixing relations are
\begin{mathletters}
\label{eqno3}
\begin{eqnarray}
        \rip{\eta} &=&  \cos \phi \rip{\eta_\NSt} - \sin \phi \rip{\eta_S} \, ,
        \label{eqno3a}
\\
        \rip{\eta'} &=& \sin \phi \rip{\eta_\NSt} + \cos \phi \rip{\eta_S}.
        \label{eqno3b}
\end{eqnarray}
\end{mathletters}
The {\it NS--S} state mixing angle $\phi$ is related
to the singlet-octet state mixing angle $\theta$ as
$ \phi = \theta + \arctan \sqrt{2} =  \theta + 54.74^\circ  \, .$
The {\it NS--S} mixing basis is more suitable for 
some quark model considerations.
In particular, in the case of our $\eta_8$ and $\eta_0$, we again point out
that Eqs.~(\ref{eta8def})--(\ref{eta0def}) do not presently define the 
octet and singlet states of the exact SU(3) flavor symmetry, but rather the 
SU(3)$_f$--inspired {\it effective} octet and singlet states, 
since $|u\bar{u}\rangle$ and $|d\bar{d}\rangle$ are practically
chiral states as opposed to a significantly heavier $|s\bar{s}\rangle$.
When the symmetry between the {\it NS} and {\it S} sectors is broken 
like this, the {\it NS--S} mixing basis is more natural in practice.
For example, if $M_P$ denotes the mass of the meson $P$ and $\alpha_{\rm em}$
the electromagnetic fine-structure constant, the expressions 
for the $\eta$ and $\eta' \to\gamma\gamma$ decay widths in this basis become 
        \begin{eqnarray}
               W(\eta\to\gamma\gamma)
        &=&
        \frac{\alpha_{\rm em}^2}{32\pi^3}
        \frac{M_\eta^3}{9f_\pi^2}
        \left[ \frac{5}{\sqrt{2}}
   \frac{f_\pi}{{\bar f}_{\pi}} \cos\phi
                -
   \frac{f_\pi}{\bar{f}_{s\bar s}} \sin\phi
        \right]^2~,
        \label{etawidth}
        \\
               W(\eta^\prime\to\gamma\gamma)
        &=&
        \frac{\alpha_{\rm em}^2}{32\pi^3}
        \frac{M_{\eta^\prime}^3}{9f_\pi^2}
        \left[ \frac{5}{\sqrt{2}}
    \frac{f_\pi}{{\bar f}_{\pi}} \sin\phi
                +
     \frac{f_\pi}{\bar{f}_{s\bar s}} \cos\phi
        \right]^2~,
        \label{etaprimewidth}
        \end{eqnarray}                                                          
instead of Eqs. (40)--(41) in Ref. \cite{KlKe2}. That is, instead of using the 
$\gamma\gamma$-decay constants ${\bar f}_{\eta_8}$ and ${\bar f}_{\eta_0}$ [see Eqs. (29)--(30) in Ref. \cite{KlKe2}] one writes
the $\gamma\gamma$ decay amplitudes through 
analogously defined $s\bar s$ two--photon decay constant
${\bar{f}_{s\bar s}}$ and the pionic $\gamma\gamma$-decay constant
${\bar f}_\pi (\approx f_\pi)$. This {\it NS--S} decomposition is more natural
for the following reasons. The ${\eta_\NSt} \to \gamma\gamma$ amplitude, 
$T_{\eta_{NS}}(0,0) = (1/\sqrt{2})[T_{u\bar u}(0,0) + T_{d\bar d}(0,0)] 
= (5/3) T_{\pi^0}(0,0)$, 
is quite close to its chiral limit value fixed by the QED axial anomaly 
[see Eqs. (26),(28) in Ref. \cite{KlKe2}] since ${\bar f}_\pi$ is approximated well 
by the usual leptonic (axial-current) decay constant $f_\pi$, while
$T_{\eta_{S}}(0,0) =  T_{s\bar s}(0,0)$
is noticeably farther from the chiral limit value.
However, at least  ${\bar f}_{s\bar s}\approx f_{s\bar s}$ is ensured through 
the Goldberger--Trieman (GT) relation (which is a natural result in SD-BS approach). 
In contradistinction to that, for the decay constants appearing in
the octet--singlet decomposition, ${\bar f}_{\eta_8} < f_{\eta_8}$
rather generally \cite{KlKe2,KeKlSc2000} in the quark--based approaches.

Another advantage of the ${\eta_\NSt}$--${\eta_S}$ state mixing angle 
$\phi$, is that one then easily notes the consistency of our (in Ref. \cite{KlKe2}
and Refs.~\cite{KeKlSc2000,Rostock2000,KeKlSc2001}) preferred mixing angle 
$\phi = 42^\circ$ 
with the value of $\phi$ obtained in the recent thorough analysis of 
Feldmann, Kroll, and Stech 
(FKS)~\cite{FeldmannKrollStech98PRD,FeldmannKrollStech99PLB}.
For reasons related to this, the {\it NS--S} mixing basis also
offers the most straightforward way to show the consistency of
our procedures and the corresponding results obtained using just one
($\theta$ or $\phi$) state mixing angle with the two-mixing-angle 
scheme \cite{Leutwyler98,KaiserLeutwyler98}, 
which is defined with respect to the mixing of the decay constants.
This is explained in detail in our Ref.~\cite{KeKlSc2000}, which
improved the analysis of mixing in the $\eta$--$\eta'$ complex over
that performed in Ref. \cite{KlKe2}, only to confirm{\footnote{See also our 
shorter Ref.~\cite{Rostock2000}.}} the preferred value of the 
state mixing angle found already in Ref. \cite{KlKe2}, namely $\phi=42^\circ$ (or 
equivalently, $\theta=-12.7^\circ$). This is practically \cite{KeKlSc2000}
the same as the result of the ``FKS scheme and theory"  
\cite{FeldmannKrollStech98PRD,FeldmannKrollStech99PLB,Feldmann99IJMPA}, 
and in agreement with data. We thus use this mixing angle value in 
        \begin{eqnarray}
        T_\eta(k^2,{k'}^2)
        &=&
 \cos\phi\, T_{\eta_{NS}}(k^2,{k'}^2) - \sin\phi\, T_{\eta_S}(k^2,{k'}^2)~,
\label{etaampl}
        \\
        T_{\eta^\prime}(k^2,{k'}^2)
        &=&
 \sin\phi\, T_{\eta_{NS}}(k^2,{k'}^2) + \cos\phi\, T_{\eta_S}(k^2,{k'}^2)~.    
\label{etaprimeampl}
\end{eqnarray}
These amplitudes are the two--photon transition form factors of $\eta$ and 
$\eta^\prime$. In Fig. 1, we follow the CELLO collaboration \cite{behrend91} in presenting 
results on the form factors in terms of the convenient combination
\begin{equation}
        \frac{\pi\alpha_{\rm em}^2 M_P^3}{4}
     \, |T_P(k^2,{k'}^2)|^2~, \qquad (P = \pi^0, \eta, \eta') \, .
        \label{generP-Widths}
        \end{equation}
Its on--shell 
limit $k^2={k'}^2=0$ returns the $\eta, \eta' \to \gamma\gamma$ widths 
(\ref{etawidth})--(\ref{etaprimewidth}) already studied in Ref. \cite{KlKe2}. 
In the 
present paper, we evaluate the amplitudes (\ref{etaampl})--(\ref{etaprimeampl})
in the cases in which one or both photons are off--shell and spacelike,
$k^2=-Q^2<0$, ${k'}^2=-{Q'}^2 \leq 0$.

In this paper 
we use the same SD-BS model~\cite{munczek92,jain93b}, model 
parameters and solutions for the dressed 
quark propagators and the corresponding quark--antiquark bound states 
as in Ref. \cite{KlKe2}. The incorporation of the quark--photon interactions
is also the same as adopted there through the scheme of a generalized 
impulse approximation, where all propagators, bound-state vertices, and
quark--photon vertices are dressed.
(This impulse approximation in the present application is illustrated by 
the pseudoscalar-photon-photon triangle graph in Fig. 1 of Ref. \cite{KlKe2}.)
This is necessary for reproducing exactly and analytically anomalous 
$\gamma\gamma$ (on--shell) amplitudes {\footnote{But also others, 
notably $\gamma \pi^+ \to \pi^+ \pi^0 $; see 
Refs. \cite{AR96,BiKl9912452}. }}    in the chiral 
limit, and requires the usage of a dressed quark--photon vertex satisfying 
the vector Ward--Takahashi identity. The Ball--Chiu (BC)~\cite{BC} vertex 
is used in Ref. \cite{KlKe2}, and thus also here. The off--shell amplitudes 
\begin{equation}
T_{\eta_{NS}}(-Q^2,-{Q'}^2) = \frac{1}{\sqrt{2}} 
       [ T_{u\bar u}(-Q^2,-{Q'}^2) + T_{d\bar d}(-Q^2,-{Q'}^2) ]
=  \frac{5}{3} T_{\pi^0}(-Q^2,-{Q'}^2)
\label{connectNSpi}
\end{equation}
(obtained by working, as in Ref. \cite{KlKe2}, in the isospin limit) and 
\begin{equation}
T_{\eta_{S}}(-Q^2,-{Q'}^2) = T_{s\bar s}(-Q^2,-{Q'}^2) \, ,
\end{equation}
are calculated numerically
in the same way as it was done for the pion in Ref. \cite{KeKl3},
also still keeping the approximation used there, which consists in 
discarding the second and higher derivatives in the momentum expansions.

\vspace{3mm}

{\bf 3.}
The results on the $\eta',\eta$, and $\pi^0$ transition form factors are 
presented in Fig. 1 in the spacelike momentum range $0 < Q^2 < 8$ GeV$^2$,
along with the experimental data \cite{behrend91,gronberg98,Acciarri+al98}.
There are three sets of three theoretical curves each. The highest
of these triplets pertains to $\eta'$, the middle one to $\eta$, 
and the lowest one to $\pi^0$. The same holds for the three distinct
groups of data. 

All of the displayed data points \cite{behrend91,gronberg98,Acciarri+al98},
as well as the solid and dotted curves in each of the curve triplets,
pertain to the $\gamma^\star\gamma$ case of one photon of spacelike 
virtuality and one real photon, $Q^2>0$ and ${Q'}^2=0$.
The solid curves in Fig. 1 represent our $\eta',\eta$, and $\pi^0$ model 
form factors (\ref{generP-Widths}), obtained through Eqs.
(\ref{etaampl})--(\ref{generP-Widths}) with our preferred $\phi=42^\circ$.

Note that all model input was fixed in Ref. \cite{jain93b} and 
Ref. \cite{KlKe2}, 
so that our transition form factors are pure predictions. The agreement 
with experiment is thus relatively good, considering the absence of any 
additional model fitting in this paper.
The main deficiency in the description of the data is that our predictions 
are too high in the intermediate range of transferred momenta, 
0.5 GeV$^2 <Q^2< 2$ GeV$^2$, at least for $\eta'$ and $\pi^0$. 

In Fig. 1, we also plot (by dotted curves) the $\eta',\eta$, and $\pi^0$  
form factors (\ref{generP-Widths}) stemming from the Brodsky--Lepage 
(BL) Ansatz $T_P^{BL}(-Q^2,0)$ for $P = \pi^0, \eta_{NS}, \eta_{S}$
\cite{BrodskyLepage81}. We do this to compare in a brief
and compact, albeit very rough way, our results with the predictions of  
Abelian axial anomaly, vector meson dominance (VMD), and perturbative
QCD (pQCD) in their regimes of validity. Namely, this Ansatz is
adjusted so that it agrees with the axial anomaly predictions at $Q^2=0$
(e.g., $T_{\pi^0}(0,0) = 1/4\pi^2 f_\pi$), while for large $Q^2$ it tends
to the behavior $\propto 1/Q^2$ predicted by pQCD \cite{BrodskyLepage81}.
Also, due to $8\pi^2 f_{\pi}^2 \approx m_\rho^2 \approx m_\omega^2$ and 
$8\pi^2 f_{\eta_{S}}^2 \approx m_\phi^2$, the BL Ansatz is not very different 
from VMD, since their corresponding residues also agree approximately 
\cite{FeldmannKroll98PRD}. The BL Ansatz was shown to work well not only 
for $P=\pi^0$, but also $P=\eta, \eta'$ \cite{FeldmannKroll98PRD}.
In the {\it NS--S} basis, it is given by 
\begin{equation}
T_P^{BL}(-Q^2,0) = \frac{N_c C_P 2 \sqrt{2} f_P}{Q^2}\,
                   \frac{1}{1 + \frac{8\pi^2 f_P^2}{Q^2}} \, ,
\, (P = \pi^0, \eta_{NS}, \eta_{S}) \, ,
\label{NS-S-BL}
\end{equation}
where $N_c = 3$
and the flavor-charge factors are $ C_{\pi^0} = 1/3\sqrt{2} $,
$ C_{\eta_{NS}} = 5/9\sqrt{2} $, and $ C_{\eta_{S}} = 1/9$.
The dotted curves stem from the BL Ans\" atze (\ref{NS-S-BL}) for the 
case named ``FKS scheme and phenomenology", which has 
$f_{\eta_{NS}}=1.07 f_\pi$, $f_{\eta_{S}}=1.34 f_\pi$ and the mixing 
angle $\phi=39.3^\circ$
\cite{FeldmannKrollStech98PRD,FeldmannKrollStech99PLB,Feldmann99IJMPA}.
``FKS scheme and theory" has $f_{\eta_{NS}}=f_\pi$ like we do, and
$f_{\eta_{S}}=1.41 f_\pi$ and $\phi=42.4^\circ$, very similar to us,
and yields curves which we do not plot because they are too close
to our predictions.

As stressed by Hayakawa and Kinoshita \cite{H+Kinoshita98},
VMD still provides one of the best fits to the $\pi^0$ transition 
form-factor data. Nevertheless, in contrast to, e.g., compliance of the 
SD-BS approach with the axial anomaly \cite{Maris:1998hd,Maris:1998hc}, 
VMD seems to be the ingredient missing in the present approach, which relies 
on the BC vertex Ansatz. Maris and Tandy \cite{MarisTandyPRC61-045202}
solved the SD equation for the dressed quark--photon vertex 
in a model similar to the present one. Their vertex solution 
exhibits the vector meson pole in the transverse part of the 
vertex, and this is the chief source of difference from the BC 
vertex Ansatz. Their vertex solution can be reasonably approximated 
for $Q^2 \geq - m_\rho^2$ by a phenomenological vertex Ansatz, the 
longitudinal part of which is given by the BC vertex, while the transverse
part contains the $\rho$-meson pole term contributing 
significantly for relatively small but nonzero $Q^2$. The transition 
form factor was calculated with the VMD-incorporating 
vertex solution only for $\pi^0$ and up to intermediate momenta, 
the $Q^2$ range where our curves overshoot. Indeed, 
the required reduction of $T_{\pi^0}(-Q^2,0)$ was found
there \cite{NT9908045}. 
{}From Eq. (\ref{connectNSpi}), it is obvious that this same mechanism would
lead, in the low and intermediate $Q^2$ range, also to the 
reduction of the $\eta$ and $\eta'$ transition form factors 
considered here.

Finally, in Fig. 1 we also plot some of our model predictions for 
the $Q^2$--dependence of the transition form factors when {\it both} 
photons are off-shell. In each curve triplet $P = \eta', \eta, \pi^0$, 
the dashed curve depicts the pertinent transition form factor 
(\ref{generP-Widths}) for the special case of 
the symmetric $\gamma^\star\gamma^\star$ virtualities, ${Q'}^2 = Q^2$. 
Such $\gamma^\star\gamma^\star$ form factors cannot be compared with 
experiment at present, since there are no published experimental data 
for any $\gamma^\star\gamma^\star \to \pi^0, \eta, \eta'$ transitions 
yet. Nevertheless, there will hopefully be some such data in the future, 
from BaBar, Belle, and CLEO \cite{Savinov:2001wj,Diehl:2001dg}.


\section*{Acknowledgments}
\noindent The authors acknowledge 
the support of the Croatian Ministry of Science and
Technology contracts No. 0119--261 and No. 009802.


\newpage

\section*{Figure captions}

\begin{itemize}

\item[{\bf Fig.~1:}] 
 
The $Q^2$ dependence of various results for the form factors of $P = \eta',
\eta,\pi^0$. The curves are obtained through Eq.  (\ref{generP-Widths}) 
employing the empirical meson masses, $M_{\eta'}=0.958$ GeV, $M_{\eta}=0.547$ 
GeV, and $M_{\pi^0}=0.135$ GeV \cite{Groom:2000in}. In the ${Q'}^2=0$ case, 
for which we plot data 
from CELLO \cite{behrend91} (circles), CLEO \cite{gronberg98} (triangles), 
and L3 \cite{Acciarri+al98} (squares), the solid curves correspond to our 
numerically obtained, model $\gamma^\star\gamma$ transition form factors, 
and the dotted curves correspond to the BL ones. The dashed curves correspond to our 
$\gamma^\star\gamma^\star$ model form factors, but for the symmetric case 
${Q'}^2=Q^2$.
\end{itemize}


\newpage

\vspace*{4cm}

\epsfxsize = 16 cm \epsfbox{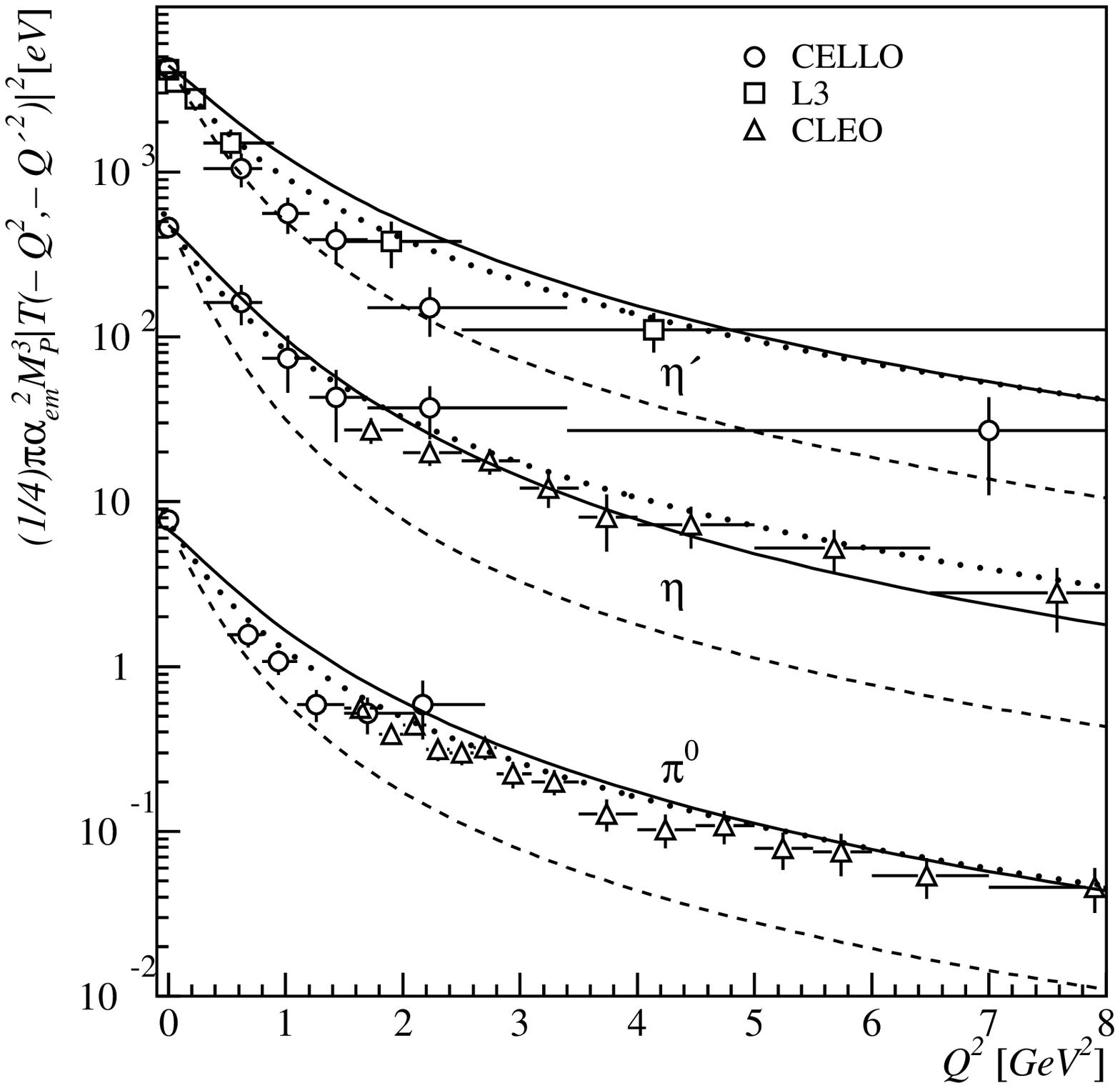}


\begin{thebibliography}{100}

\bibitem{KlKe2}
D. Klabu\v car and D. Kekez, {Phys. Rev.} D {\bf 58}, 096003 (1998).             
\bibitem{RW}
C. D. Roberts and A. G. Williams,
Prog. Part. Nucl. Phys. {\bf{33}}, 477 (1994).
                                                                                
\bibitem{AlkoferSmekal00}
R. Alkofer and L. von Smekal,
Phys. Rep. {\bf 353}, 281 (2001).


\bibitem{RobertsSchmidt00}
C. D. Roberts and S. M. Schmidt,
Prog. Part. Nucl. Phys. {\bf 45}, S1 (2000).
                                                                                
\bibitem{KeKl1}
D. Kekez and D. Klabu\v car, {Phys. Lett.} B {\bf 387}, 14 (1996).              

\bibitem{KeBiKl98}
D. Kekez, B. Bistrovi\' c, and D. Klabu\v car, 
Int. J. Mod. Phys. A {\bf 14}, 161 (1999).

\bibitem{KeKl3}
D. Kekez and D. Klabu\v car, {Phys. Lett.} B {\bf 457}, 359 (1999).             

\bibitem{KlKe4}
D. Klabu\v car and D. Kekez, Fizika B {\bf 8}, 303 (1999).             

\bibitem{munczek92}
H. J. Munczek and P. Jain, Phys. Rev. D {\bf{46}}, 438 (1992).
 
\bibitem{jain93b}
P. Jain and H. J. Munczek, Phys. Rev. D {\bf{48}}, 5403 (1993).           

\bibitem{KeKlSc2000}
D. Kekez, D. Klabu\v car, and M. Scadron,
J. Phys. G {\bf 26}, 1335 (2000).

\bibitem{Rostock2000}
D. Klabu\v car, D. Kekez, and M. Scadron:
``On the $\eta$--$\eta^\prime$ complex in the SD--BS approach",
p. 201 in ``Exploring Quark Matter", proceedings of the
Workshop on Quark matter in Astro- and Particle Physics,
Rostock, Germany, 27.--29. Nov. 2000, Editors: G. R. G. Burau,
D. B. Blaschke, and S. M. Schmidt.
(Also available as hep-ph/0012267.)
 
\bibitem{KeKlSc2001}
D. Kekez, D. Klabu\v car, and M. Scadron,
J. Phys. G {\bf 27}, 1775 (2001).
                                                                                
 
\bibitem{FeldmannKrollStech98PRD}
Th. Feldmann, P. Kroll, and B. Stech, Phys. Rev. D {\bf 58}, 114006 (1998).

\bibitem{FeldmannKrollStech99PLB}
Th. Feldmann, P. Kroll, and B. Stech, Phys. Lett. B {\bf 449}, 339 (1999).

\bibitem{Leutwyler98}
H. Leutwyler, Nucl. Phys. B (Proc. Suppl.) {\bf 64}, 223 (1998).

\bibitem{KaiserLeutwyler98}
R. Kaiser and H. Leutwyler, hep-ph/9806336.

\bibitem{Feldmann99IJMPA}
See also Table 2 of the review 
Th. Feldmann, Int. J. Mod. Phys. A {\bf 15}, 159 (2000).

\bibitem{behrend91}
H.-J. Behrend {\it et al.} [CELLO collaboration],
Z. Phys. C {\bf 49}, 401 (1991).
 
\bibitem{AR96}
R. Alkofer and C. D. Roberts, Phys. Lett. B {\bf{369}}, 101 (1996).

\bibitem{BiKl9912452}
B. Bistrovi\'{c} and D. Klabu\v{c}ar, 
Phys. Lett. B {\bf 478}, 127 (2000).

\bibitem{BC}
J. S. Ball and T.-W. Chiu, Phys. Rev. D {\bf 22}, 2542 (1980).

\bibitem{gronberg98}
J. Gronberg {\it et al.} [CLEO collaboration],
Phys. Rev. D {\bf 57}, 33 (1998).
 
\bibitem{Acciarri+al98}
M. Acciarri {\em et al.} [L3 collaboration],
Phys. Lett. B {\bf 418}, 399 (1998).                                            

\bibitem{BrodskyLepage81}
S. J. Brodsky and G. P. Lepage, Phys. Rev. D {\bf 24}, 1808 (1981).             

\bibitem{FeldmannKroll98PRD}
Th. Feldmann and P. Kroll, Phys. Rev. D {\bf 58}, 057501 (1998).


\bibitem{H+Kinoshita98}
M. Hayakawa and T. Kinoshita, Phys. Rev. D {\bf 57}, 465 (1998).


\bibitem{Maris:1998hd}
P. Maris, C. D. Roberts, and P. C. Tandy,
Phys.\ Lett.\ B {\bf 420}, 267 (1998).

\bibitem{Maris:1998hc}
P. Maris and C. D. Roberts,
Phys.\ Rev.\ C {\bf 58}, 3659 (1998).

\bibitem{MarisTandyPRC61-045202}
P. Maris and P. C. Tandy, Phys. Rev. C {\bf 61}, 045202 (2000). 

\bibitem{NT9908045}
P. Maris and P. C. Tandy, Nucl. Phys. A {\bf 663}, 401 (2000).

\bibitem{Savinov:2001wj}
V. Savinov,
hep-ex/0106013.
 
\bibitem{Diehl:2001dg}
M. Diehl, P. Kroll, and C. Vogt,
hep-ph/0108220.

\bibitem{Groom:2000in}
D.~E.~Groom {\it et al.}  [Particle Data Group Collaboration],
Eur.\ Phys.\ J.\ C {\bf 15}, 1 (2000).

\end{thebibliography}
\end{document}